\documentclass[amstex,12pt]{article}
\usepackage{a4,isolatin1}
\usepackage{oldgerm}
\usepackage{amsthm}
\usepackage{amsmath, mathrsfs, amssymb,amscd}

\font\fa=bbm11
\newcommand{\R}{\mbox{\fa R}}   
\newcommand{\N}{\mbox{\fa N}}   

\newcommand{\Ibb}[1]{ {\rm I\ifmmode\mkern -3.6mu\else\kern -.2em\fi#1}}
\newcommand{\ibb}[1]{\leavevmode\hbox{\kern.3em\vrule
     height 1.2ex depth -.3ex width .2pt\kern-.3em\rm#1}}
\newcommand{\Cl}{{\ibb C}}           
\newcommand{\Rl}{{\Ibb R}}           

\newtheorem{definition}{Definition}
\newtheorem{lemma}{Lemma}
\newtheorem{proposition}{Proposition}

\newcommand{\Hil}{\mathcal{H}}
\newcommand{\DD}{\mathcal{D}} 
\newcommand{\W}{\mathcal{W}}    
\newcommand{\Z}{\mathcal{Z}}    
\newcommand{\Ss}{\mathscr{S}}   
\newcommand{\OO}{\mathcal{O}}   
\newcommand{\PGpo}{\mathcal{P}_+^\uparrow}   

\newcommand{\frS}{\textfrak{S}} 

\newcommand{\supp}{\mathrm{supp}}

\newcommand{\te}{\theta}
\newcommand{\la}{\lambda}

\newcommand{\fti}{\widetilde{f}}
\newcommand{\gti}{\widetilde{g}}
\newcommand{\fbar}{\overline{f}}

\newcommand{\Zd}{Z^{\dagger}}
\newcommand{\zd}{z^{\dagger}}

\newcommand{\ad}{a^{\dagger}}

\newcommand{\iin}{_\mathrm{in}}
\newcommand{\oout}{_\mathrm{out}}
\newcommand{\tp}[1]{^{\otimes #1}}    

\begin{document}

\title{Polarization-Free Quantum Fields and Interaction}
\author{Gandalf Lechner\\Institut für Theoretische Physik, Universität Göttingen\\37073 Göttingen,
  Germany\\{\small e-mail:{\tt lechner@theorie.physik.uni-goettingen.de}}}
\maketitle
\abstract{A new approach to the inverse scattering problem proposed by
  Schroer, is applied to two-dimensional integrable quantum field
  theories. For any two-particle S-matrix $S_2$ which is analytic in
  the physical sheet, quantum fields are constructed which are
  localizable in wedge-shaped regions of Minkowski space and whose
  two-particle scattering is described by the given $S_2$. These
  fields are polarization-free in the sense that they create one-particle states from the vacuum without polarization clouds. Thus they provide examples of temperate polarization-free generators in the presence of non-trivial interaction.}

\section{Introduction}

Within the context of collision theory the inverse scattering
problem, i.e. the reconstruction of a relativistic quantum field
theory from its scattering matrix, is one of a few remaining challenges. This
question is important for the construction of field-theoretic models
and is addressed in the framework of integrable theories in the so-called form-factor program \cite{ff}.\\
It was realized by Schroer \cite{S97} that in certain models of this
type an
intermediate step from the S-matrix to the
local quantum fields might be possible. These models admit an algebra of particle creation and annihilation operators satisfying exchange relations involving the two-particle S-matrix \cite{zamo}. Schroer argued that these operators can be combined to yield Bose-type fields which are localized in wedge-shaped regions and which create single particle states from the vacuum without accompanying polarization clouds ({\bf p}olarization-{\bf f}ree {\bf g}enerators). Because of their relations to the S-matrix, they promised to provide a
new tool in the inverse scattering problem in quantum field theory.\\
A model-independent analysis of these new objects was carried out in
\cite{BBS}. It turned out that polarization-free generators exist in
any local theory, but their subtle domain properties render them useless, in general. Assuming for example certain mild temperateness conditions, which are inevitable if one wants to do scattering theory, it was shown in \cite{BBS} that the underlying S-matrix must be trivial if the space-time dimension is larger than two. The results obtained for the
two-dimensional case indicated that temperate PFGs might well exist in
theories with a factorizing S-matrix.\\
Based on the ideas by Schroer \cite{S97, schroer}, we prove in this
article by an explicit construction, cf. also \cite{diplom}, that there are indeed two-dimensional,
factorizing theories with non-trivial interaction which allow for temperate PFGs localized in wedge regions. For notational
simplicity, we consider only one type of massive, scalar particles. Given a two-particle S-matrix $S_2$ which has to
satisfy well-known physical requirements and additionally
analyticity in the physical sheet, we construct temperate polarization-free
generators as semi-local Wightman fields which give rise to a
wedge-local theory whose two-particle scattering is described by $S_2$.\\
Section 2 introduces Zamolodchikov's algebra which is the basis for
our construction and a representation of this algebra on a Hilbert
space similar to the bosonic Fock space. In section 3 we define and
analyze the wedge-local field. Collision processes of two particles
are studied in section 4.

\section{The Zamolodchikov algebra}

The starting point of our construction is the well-known
Zamolodchikov-Faddeev algebra \cite{zamo} which is commonly used in
the context of two-dimensional, integrable models \cite{abda}. This
algebra is described by quadratic exchange relations involving an
operator $S_2$ which is closely related to the two-particle S-matrix. Since we are dealing with the inverse
scattering problem, $S_2$ is assumed to be given.\\
We define the Zamoldchikov algebra in a somewhat abstract manner: Let
$H$ be a Hilbert space with an antiunitary involution 
$J_0=J_0^{-1}=J_0^*$
acting on it and let $S_2:H\otimes H\to H\otimes H$ be a unitary
operator which fulfills the following exchange relations with $J_0$
and the ``flip operator'' $t:H\otimes H\to H \otimes H$,
$t(\psi\otimes\varphi):=\varphi\otimes\psi$:
\begin{eqnarray}\label{scon}
  S_2 t = tS_2^*,\qquad\qquad S_2J_0\tp{2}=J_0\tp{2}S_2^*\;.
\end{eqnarray}
Here and in the following $A\tp{n}$ denotes the $n$-fold tensor
product of an operator $A$ on $H$, and $H\tp{n}$ is the $n$-fold
tensor product of $H$. The constraints (\ref{scon}) will translate to
physically significant properties of the two-particle S-matrix in the
representation of the Zamolodchikov algebra given below.\\
Consider the free $*$-algebra generated
by the symbols $1$, $Z(\psi),\Zd(\psi)$, $\psi\in H$ and
$(Z\times Z)(\psi^{(2)})$, $(\Zd\times\Zd)(\psi^{(2)})$,
$(Z\times\Zd)(\psi^{(2)})$, $(\Zd\times Z)(\psi^{(2)})$,
$\psi^{(2)}\in H\tp{2}$. The Zamolodchikov algebra $\Z(H,J_0,S_2)$ is
obtained by imposing several additional relations on this free
algebra: All symbols are assumed to depend complex linearly on
$\psi\in H$ and $\psi^{(2)}\in H\tp{2}$, respectively. With respect to
multiplication in $\Z$ we have $1$ as the neutral element and
\begin{eqnarray}\label{z-mult}
  \Zd(\psi)Z(\varphi) = (\Zd\times Z)(\psi\otimes\varphi),\qquad\quad\psi,\varphi\in H,
\end{eqnarray}
and analogously for the other symbols. The star operation on $\Z$ is
fixed by
\begin{eqnarray}\label{star}
  1^* = 1,\qquad Z(\psi)^* = \Zd(J_0\psi),\qquad\psi\in H\,.
\end{eqnarray}
The definition of the algebraic structure of $\Z$ is completed by
requiring the relations ($\varphi,\psi\in H$)
\begin{eqnarray}
  \Zd(\varphi)\Zd(\psi) &=& (\Zd\times 
\Zd)(S_2^*(\psi\otimes\varphi))\label{z1}\;,\\
  Z(\varphi)\Zd(\psi) &=& (\Zd\times
  Z)(S_2(\psi\otimes\varphi)) + \langle J_0\varphi,\psi\rangle
  \cdot 1\;. \label{z2}
\end{eqnarray}
Here $\langle.\,,\,.\rangle$ denotes the scalar product in $H$.
It should be mentioned that under the assumptions  made the exchange relation
$Z(\varphi)Z(\psi)=(Z\times Z)(S_2^*(\psi\otimes\varphi))$ holds true as
well. All these equalities are consistent because of the properties
(\ref{scon}) of $S_2$.\\ 
\\
Having defined the abstract Zamolodchikov algebra, we now turn to the
construction of a field theoretic model based on a specific
representation of $\Z$. For the sake of simplicity we consider here
only the case of one type of scalar particles of mass $m$, although
our results apply to a more complicated particle spectrum as well.\\
It is convenient to parametrize the momentum $p$ of a particle by the rapidity $\te$,
\begin{equation}
 p(\te) = m\left(
   \begin{array}{c}
     \cosh\te\\\sinh\te
   \end{array}
   \right)\;.
\end{equation}
We choose as our Hilbert space the rapidity space of a particle of
mass $m$, $H:=\Hil_1:=L^2(\Rl,\mathrm{d}\te)$. The inversion $J_0$ is defined as complex conjugation in rapidity space,
\begin{equation}
  (J_0\psi)(\te) := \overline{\psi(\te)},\qquad \psi\in L^2(\Rl,\mathrm{d}\te)\,.
\end{equation}
To specify the operator $S_2$, we consider a function (denoted by $S_2$
as well) which is meromorphic on the strip $S(0,\pi)=\{\zeta\in\Cl\,:\,0<\mathrm{Im}(\zeta)<\pi\}$ in the complex
rapidity plane and continuous on its closure. Furthermore, $S_2$
should satisfy the following equations for $\te\in \Rl$:
\begin{eqnarray}
  S_2(\te)^{-1} = \overline{S_2(\te)} =  S_2(-\te) = S_2(\te+i\pi)\,.\label{s_constraints}
\end{eqnarray}
The function $S_2$ well be called ``scattering function'' in the
following, and the operator $S_2$ acts by multiplication with it:
\begin{eqnarray}
  (S_2\psi^{(2)})(\te_1,\te_2) :=
  S_2(\te_1-\te_2)\cdot\psi^{(2)}(\te_1,\te_2),\qquad
  \psi^{(2)}\in\Hil_1\tp{2}\,.
\end{eqnarray}
The above stated properties of the scattering function imply that
$S_2$ is a unitary operator on $L^2(\Rl^2)$ which fulfills the
consistency requirements
(\ref{scon}) of the Zamolodchikov algebra. The constraints
(\ref{s_constraints}) are the well-known properties of unitarity, real
analyticity and crossing symmetry (See \cite[chapter 3]{korff} for a
review) which are known to be satisfied by any two-particle S-matrix
element obtained from an integrable local quantum field theory on
two-dimensional Minkowski space.\\
Later on we will
have to restrict attention to functions $S_2$ which are bounded on the
closed strip $\overline{S(0,\pi)}$ and analytic on its interior, but for the moment the constraints summarized in
(\ref{s_constraints}) are sufficient to proceed.\\
After these preparations we can define our representation space; it
will be a proper supspace of $\mathcal{F}_{\Hil_1}:=\bigoplus_{n=0}^\infty\Hil_1\tp{n}$. The
construction is very similar to the construction of the free bosonic
or fermionic Fock spaces, the difference being a different
symmetrization procedure which depends on the choice of $S_2$. In a
different context, this construction has already been carried out in
\cite{li_min}.\\
In the lemma stated below and the subsequent computations we use the
following notation: Given an operator $A$ acting on $\Hil_1\otimes \Hil_1$ we denote by $A_n^{i,j}$ the operator
which acts on the $n$-fold tensor product space $\Hil_1\tp{n}$ like $A$ on the $i^{\mathrm{th}}$ and $j^{\mathrm{th}}$ factor and trivially
on the remaining factors, e.g. $A\otimes 1_n = A_{n+2}^{1,2}$
with $1_n$ denoting the identity on $\Hil_1\tp{n}$.\\
Furthermore, $\frS_n$ denotes the group of permutations of $n$
elements, and $\tau_i\in\frS_n$, $i=1,...,n-1$, the transposition which
exchanges $i$ and $(i+1)$. As in \cite{li_min}, we have the following lemma:
\begin{lemma}
  Let $R :\Hil_1\tp{2}\to\Hil_1\tp{2},R(\psi\otimes\varphi):=
  S_2^*(\varphi\otimes\psi)$. The map 
  \begin{equation}
    \tau_i \longmapsto D_n(\tau_i) := R_n^{i,i+1} 
  \end{equation}
  defines a unitary representation of $\frS_n$ on
  $\Hil_1\tp{n}$. Moreover, the mean
  \begin{equation}
    P_n := \frac{1}{n!}\sum_{\pi\in\frS_n}D_n(\pi)\,,
  \end{equation}
is an orthogonal projection.
\end{lemma}
\begin{proof}
  In order to show that $D_n$ is a representation of $\frS_n$, one has
  to check the relations $D_n(\tau_i)^2=1$ for $i=1,...,n-1$ and
  $[D_n(\tau_i),D_n(\tau_j)]=0$ for $|i-j|>1$, as well as
\begin{eqnarray}\label{YB}
  D_n(\tau_i)D_n(\tau_{i+1})D_n(\tau_i)=D_n(\tau_{i+1})D_n(\tau_i)D_n(\tau_{i+1})\;\qquad i=1,...,n-2\;.
\end{eqnarray}
  The constraints (\ref{s_constraints}) imply the first relation, and
  the second equality holds because $D_n(\tau_i)$ acts only on the tensor factors $i$ and $(i+1)$. Since
  $S_2^*$ is a multiplication operator, the different $S_2$-factors occuring in the computation of $D_n(\tau_i)D_n(\tau_{i+1})D_n(\tau_i)$ and
  $D_n(\tau_{i+1})D_n(\tau_i)D_n(\tau_{i+1})$ commute with each other,
  which implies the third relation.\\
The unitarity of $R$ follows from (\ref{s_constraints}) and carries
over to $D_n(\pi)$ since the transpositions generate $\frS_n$. In
particular, the unitarity of the representation $D_n$ implies
$P_n=P_n^*$. The equation $P_n^2=P_n$ holds because $\frS_n$ is a
group of $n!$ elements.
\end{proof}

With the help of the projections $P_n$ we can symmetrize the
unsymmetrized Fock space $\mathcal{F}_{\Hil_1}$ with respect to $D_n$ which
yields the definition of our Hilbert space $\Hil$:
\begin{eqnarray}
  \Hil := \bigoplus_{n=0}^\infty \Hil_n,\qquad \Hil_n := P_n\Hil_1\tp{n},\qquad\Hil_0 := \Cl\cdot\Omega\,.
\end{eqnarray}
The projections $P_n$ will be extended to operators on
$\mathcal{F}_{\Hil_1}$ by $P_n\psi:=0$ for $\psi\notin \Hil_1\tp{n}$,
and we introduce the notation $P:=\oplus_{n=0}^\infty P_n$.\\
The zero-particle vector $\Omega=(1,0,0,...)\in\Hil$ is interpreted 
as
the physical vacuum. Generalizing the totally symmetric functions
known from the free bosonic field, vectors $\psi^{(n)}\in\Hil_n$ are functions in
$L^2(\Rl^n)$ with the property
\begin{equation*}
  \psi^{(n)}(\te_1,...,\te_{i+1},\te_i,...\te_n) = (S_2
  \psi^{(n)})(\te_1,...,\te_n) = S_2(\te_i-\te_{i+1})\psi^{(n)}(\te_1,...,\te_n)\,.
\end{equation*}

As in the free case, we have a representation of the proper orthochronous
Poincar\'e group $\PGpo$ on $\Hil$. In $1+1$ dimensions $\PGpo$ is
generated by the translations $x\longmapsto x+a$ and the velocity
transformations (``boosts'')
\begin{equation*}
  x\longmapsto\left(
    \begin{array}{cc}
      \cosh\la & \sinh\la\\
      \sinh\la & \cosh\la
    \end{array}
  \right)x\;.
\end{equation*}
The Poincar\'e transformations will be denoted
by $(a,\la)$, where $a\in\Rl^2$ is a space-time translation and
$\la\in\Rl$ the rapidity defining the boost. For $(a,\la)\in\PGpo$ we set $U_1(a,\la):\Hil_1\to\Hil_1$,
  \begin{equation}\label{rep_U}
    (U_1(a,\la)\psi)(\te) := e^{ip(\te)a}\cdot\psi(\te-\la)
  \end{equation}
  and define $U(a,\la)$ on $\mathcal{F}_{\Hil_1}$ via second quantization, $U(a,\la) := 
\bigoplus_{n=0}^\infty U_1(a,\la)\tp{n}$. We also use the shorter 
notation $U(a):=U(a,0)$.\\
Since the $S_2$-factors occuring in the symmetrization $P$ depend only
on differences of rapidities and since $U(a)$ and $S_2^*$ are both
multiplication operators, $P$ and $U(a,\la)$ commute. Thus we conclude
that $U(a,\la)$ is also a well defined, unitary operator on the
subspace $\Hil\subset\mathcal{F}_{\Hil_1}$. In view of the spectral properties of $U_1(a,0)$, $U$
is a positive energy representation of $\PGpo$.\\

Guided by the Fock representation of the CCR algebra, we now introduce properly symmetrized creation and annihilation operators on $\Hil$. We start from their unsymmetrized counterparts $a(\varphi),\ad(\varphi)$, $\varphi\in\Hil_1$, which are given on $\psi_1\otimes...\otimes\psi_n\in\Hil_1\tp{n}$ by the formulae
\begin{eqnarray}
   a(\varphi)\Omega &:=& 0\;,\quad
  \;\;a(\varphi)\psi_1\otimes ...\otimes\psi_n  \;:=\; \sqrt{n}\,\langle J_0\varphi,\psi_1\rangle\,\psi_2\otimes...\otimes\psi_n,\label{a3}\\
\ad(\varphi)\Omega &:=&  \varphi\;,\quad
\ad(\varphi)\psi_1\otimes ...\otimes\psi_n  \;:=\; \sqrt{n+1}\;\varphi\otimes\psi_1\otimes...\otimes\psi_n\;. \label{a4} 
\end{eqnarray}
These expressions may be extended linearly to unbounded operators
defined on the subspace $\mathcal{F}_{\Hil_1}^{(0)}\subset\mathcal{F}_{\Hil_1}$ of finite particle
number. Contrary
to the common definition here $a(\varphi)$ and $\ad(\varphi)$ both
depend linearly on $\varphi$. Therefore these operators are related by
$a(\varphi)^* \supset \ad(J_0\varphi)$.\\
The domain of the representation of the Zamolodchikov algebra will be
the dense subspace $\DD := P\mathcal{F}_{\Hil_1}^{(0)}$ of symmetrized vectors of finite particle number.
\begin{lemma}\label{z-lemma}
  The map 
  \begin{equation}
  Z(\varphi) \longmapsto z(\varphi):=Pa(\varphi)P,\qquad \Zd(\varphi) \longmapsto \zd(\varphi):=P\ad(\varphi)P    
  \end{equation}
  extends to a representation of the Zamolodchikov
  algebra $\Z(\Hil_1,J_0,S_2)$ on $\DD$ for which $\Omega$ is a
  cyclic vector. On $\DD$ one has
  \begin{equation}
    \zd(\varphi) = P\ad(\varphi),\quad\qquad z(\varphi) = a(\varphi)\;.
  \end{equation}
\end{lemma}
\begin{proof}
  Since $\|P\|=1$, $z(\varphi)$ and $\zd(\varphi)$ are well-defined operators on $\DD$. Obviously $\varphi\mapsto z(\varphi)$, $\varphi\mapsto\zd(\varphi)$ are linear. For the contraction caused by the annihilation operator we will use
  the notation 
  \begin{eqnarray}
    I_n^{i,j}[\psi_1\otimes...\otimes\psi_n] := 
    \langle J_0 \psi_i,\psi_j\rangle\,
    \psi_1\otimes...\otimes \widehat{\psi}_i \otimes ...\otimes \widehat{\psi}_j\otimes...\otimes\psi_n\,,
  \end{eqnarray}
  where the hat on
  $\widehat{\psi}_i$ denotes omission of the corresponding tensor factor. Thus 
  \begin{equation*}
    a(\varphi)\psi^{(n)} = \sqrt{n}I_{n+1}^{1,2}[\varphi\otimes \psi^{(n)}]\,,\qquad\psi^{(n)}\in\Hil_1\tp{n}\,.
  \end{equation*}
  Any symmetrization in the last $(n-1)$ tensor factors of
  $\psi^{(n)}$ remains unchanged when $a(\varphi)$ is
  applied, i.e. $a(\varphi)(1_1\otimes
  D_{n-1}(\pi))\psi^{(n)}=D_{n-1}(\pi)a(\varphi)\psi^{(n)}$ for any
  $\pi\in\frS_{n-1}$. This shows that $a(\varphi)\Hil_n \subset
  \Hil_{n-1}$. Therefore the projections $P$ in the definition of
  $z(\varphi)$ may be omitted: $z(\varphi)\psi = a(\varphi)\psi$ for all $\psi\in\DD$. Since $P=1$ on $\Hil$, the equation $\zd(\varphi)\psi = P\ad(\varphi)\psi$ follows as well.\\
 The $*$-operation acts like
\begin{equation*}
  z(\varphi)^* = (Pa(\varphi)P)^* \supset P \ad(J_0\varphi)P = \zd(J_0\varphi)
\end{equation*}
in agreement with (\ref{star}). To check the relations
(\ref{z1}, \ref{z2}), let $\psi^{(n)} \in\Hil_n,\varphi_1,\varphi_2\in\Hil_1$. Using $D_{n+2}(\tau_1)^2=1$ and $P_{n+2}D_{n+2}(\tau_1)=P_{n+2}$, we get
\begin{eqnarray*}
  \zd(\varphi_1)\zd(\varphi_2)\psi^{(n)} &=& \sqrt{(n+1)(n+2)}P_{n+2}(\varphi_1 \otimes P_{n+1}(\varphi_2 \otimes \psi^{(n)}))\\
 &=&\sqrt{(n+1)(n+2)}P_{n+2}D_{n+2}(\tau_1)(S_2^*(\varphi_2 \otimes \varphi_1) \otimes \psi^{(n)})\\
 &=& (\zd\times\zd)(S_2^*(\varphi_2\otimes\varphi_1))\psi^{(n)}\,.
\end{eqnarray*}
Thus the commutation relation (\ref{z1}) is verified, and since for
$\varphi^{(2)}\in\Hil_1\tp{2}$ we have \\$(\zd\times\zd)(\varphi^{(2)})\psi^{(n)}=\sqrt{(n+1)(n+2)}P_{n+2}(\varphi^{(2)}\otimes\psi^{(n)})$ and $(z\times z)(\varphi^{(2)})\psi^{n}=\\\sqrt{n(n-1)}I_n^{1,2}I_{n+2}^{2,3}[\varphi^{(2)}\otimes\psi^{(n)}]$, we get the correct multiplication structure (\ref{z-mult}).\\
  For the second
relation (\ref{z2}) we consider the special permutations
$\sigma_k\in\frS_n$, $k=1,...,n$ defined by
\begin{eqnarray}\label{sigma}
  \sigma_k := \tau_{k-1}\cdots\tau_1,\qquad k=2,...,n,\qquad\sigma_1 := \mathrm{id}\;.
\end{eqnarray}
In the following we will denote by $D_n^+$ the ``bosonic''
representation of $\frS_n$ which is given as the special case
$S_2\equiv 1$ of $D_n$. Using
$D_n^+(\tau_l)S_n^{i,j}=S_n^{\tau_l(i),\tau_l(j)}D_n^+(\tau_l)$, we
compute the unitaries representing $\sigma_k$:
\begin{eqnarray}\label{d_sigma}
  D_n(\sigma_k) =
  S_n^{k,k-1}D_n^+(\tau_{k-1})\cdots
  S_n^{2,1}D_n^+(\tau_1) = \prod_{j=1}^{k-1}S_n^{k,j}\;D_n^+(\sigma_k)\;.
\end{eqnarray}
Any permutation $\pi\in\frS_{n+1}$ may be decomposed according to
$\pi=\sigma_k\rho$, where $\rho\in\frS_{n+1}$ and $k\in\{1,...,n+1\}$ are uniquely determined by the condition $\rho(1)=1$. For the projection $P_{n+1}$ this entails
\begin{eqnarray*}
  P_{n+1}=\frac{1}{(n+1)!}\sum_{k=1}^{n+1}\sum_{\rho\in\frS_n}D_{n+1}(\sigma_k)(1_1\otimes D_n(\rho)) = \frac{1}{n+1}\sum_{k=1}^{n+1}D_{n+1}(\sigma_k)\;(1_1\otimes P_n)\;.
\end{eqnarray*}
With this formula at hand we can verify the second commutation
relation:
\begin{eqnarray}
  z(\varphi_1)\zd(\varphi_2)\psi^{(n)} &=&
  (n+1)I_{n+2}^{1,2}[\varphi_1 \otimes
  P_{n+1}(\varphi_2\otimes\psi^{(n)})]\nonumber\\
  &=&
  \sum_{k=1}^{n+1}I_{n+2}^{1,2}[\varphi_1\otimes\prod_{j=1}^{k-1}S_{n+1}^{k,j}D_{n+1}^+(\sigma_k)(\varphi_2\otimes\psi^{(n)})]\label{2commrelpr}
\end{eqnarray}
The term corresponding to $k=1$ gives the contribution
$\langle J_0\varphi_1,\varphi_2\rangle \psi^{(n)}$ as expected
from (\ref{z2}). We evaluate the remaining sum at
$(\te_1,...,\te_n)\in\R^n$ and get
\begin{eqnarray}\label{alg}
  \sum_{k=1}^n \int\mathrm{d}\te_0\,\varphi_1(\te_0)\varphi_2(\te_k)\prod_{j=0}^{k-1}S_2(\te_k-\te_j)\psi^{(n)}(\te_0,\te_1,...,\widehat{\te_k},...,\te_n)\;,
\end{eqnarray}
where the hat indicates omission of the variable $\te_k$.
We now compute the action of $\zd(\varphi_2)z(\varphi_1)$ on $\psi^{(n)}$:
\begin{eqnarray}
  \zd(\varphi_2)z(\varphi_1)\psi^{(n)} &=& nP_n(\varphi_2\otimes
  I_{n+1}^{1,2}[\varphi_1\otimes \psi^{(n)}])\nonumber\\
  &=& \sum_{k=1}^n \prod_{j=1}^{k-1}S_n^{k,j} D_n^+(\sigma_k)I_{n+2}^{2,3}[\varphi_2\otimes\varphi_1\otimes\psi^{(n)}]\label{2commrelpr2}\;.
\end{eqnarray}
Linear and continuous extension of (\ref{2commrelpr}) in
$\varphi_1\otimes\varphi_2$ and (\ref{2commrelpr2}) in
$\varphi_2\otimes \varphi_1$ yields the operators
$(z\times\zd)(\varphi^{(2)})$, $(\zd\times z)(\varphi^{(2)})$,
$\varphi^{(2)}\in\Hil_1\tp{2}$, respectively. In particular we have
\begin{eqnarray*}
  (\zd\times z)(S_2(\varphi_2\otimes \varphi_1))\psi^{(n)} = \sum_{k=1}^n \prod_{j=1}^{k-1}S_n^{k,j} D_n^+(\sigma_k)I_{n+2}^{2,3}[S_2(\varphi_2\otimes\varphi_1)\otimes\psi^{(n)}]\;,
\end{eqnarray*}
and an evaluation at $(\te_1,...\te_n)$ yields
\begin{eqnarray*}
  &&\sum_{k=1}^n \prod_{j=1}^{k-1} S_2(\te_k-\te_j)\big(I_{n+2}^{2,3}[S_2(\varphi_2\otimes\varphi_1)\otimes\psi^{(n)}]\big)(\te_k,\te_1,...,\widehat{\te_k},...,\te_n)\\
&=&\sum_{k=1}^n \prod_{j=1}^{k-1} S_2(\te_k-\te_j)\int\mathrm{d}\te_0 S_2(\te_k-\te_0)\varphi_2(\te_k)\varphi_1(\te_0)\psi^{(n)}(\te_0,\te_1,...,\widehat{\te_k},...,\te_n)
\end{eqnarray*}
in agreement with (\ref{alg}). So we have established (\ref{z-mult}) and the desired relation 
$z(\varphi_1)\zd(\varphi_2)\psi = (\zd\times
z)(S_2(\varphi_2\otimes\varphi_1)) \psi + \langle
J_0\varphi_1,\varphi_2\rangle\,\psi$ for all $\psi\in\DD$.\\
Finally, the cyclicity of $\Omega$ follows from
$\zd(\psi_1)\cdots\zd(\psi_n)\Omega =
\sqrt{n!}P_n(\psi_1\otimes...\otimes\psi_n)$ since $P_n$ is linear and
continuous.
\end{proof}
Usually the Zamolodchikov algebra is described in terms of
operator-valued distributions $z(\te)$, $\zd(\te)$ \cite{S97,zamo,schroer,abda,li_min}. With
this notation, our Zamolodchikov operators arise as the formal integrals $\zd(\psi)=\int\mathrm{d}\te\,\psi(\te)\zd(\te)$ and
$z(\varphi)=\int\mathrm{d}\te\,\varphi(\te) z(\te)$.\\
In the special cases $S_2 \equiv 1$ and $S_2\equiv -1$ the
Zamolodchikov algebra turns into the CCR and CAR algebra,
respectively. Subsequently we will construct bosonic quantum fields
for every function $S_2$ fulfilling (\ref{s_constraints}), such that
our construction agrees with the usual free field in the case
$S_2\equiv 1$.
\newpage

\section{Wedge-local Fields}
\setcounter{equation}{0}

With the help of the Zamoldchikov operators, we now define a quantum
field $\Phi$ on two-dimensional Minkowski space $\Rl^2$. In the following, $\Ss$ denotes the space of Schwartz test functions.
\begin{definition}
  Let $f\in\Ss(\R^2)$ 
 and set
  \begin{equation}\label{Phi}
    f^\pm(\te) := \frac{1}{2\pi}\int\mathrm{d}^2x\,f(\pm x)e^{ip(\te)x},\qquad p(\te)=m\left(
      \begin{array}{c}
        \cosh\te\\
        \sinh\te
      \end{array}
\right)\;.
  \end{equation}
  We regard $f^\pm$ as elements of $\Hil_1$. 
  The field operator $\Phi(f)$ is defined as 
  \begin{equation}
    \Phi(f) := \zd(f^+)+z(f^-)\;.
  \end{equation}
\end{definition}
In the following proposition we show that $\Phi$ shares many
properties with the free Bose field, except locality. 

\begin{proposition}
  The field operator $\Phi(f)$ has the following properties:
  \begin{enumerate}
  \item $\Phi(f)$ is an unbounded operator defined on $\DD$ which leaves this space invariant.
  \item For $\psi\in\DD$ one has
    \begin{equation}
      \Phi(f)^*\psi = \Phi(\fbar)\psi.
    \end{equation}
    All vectors in $\DD$ are entire analytic for
    $\Phi(f)$. If $f\in\Ss(\R^2)$ is real, $\Phi(f)$ is
    essentially self-adjoint on $\DD$.
    \item $\Phi$ is a solution of the Klein-Gordon equation: For every
      $f\in\Ss(\Rl^2)$, $\psi\in\DD$ one has 
      \begin{eqnarray}
        \Phi((\Box+m^2)f)\psi = 0\,.
      \end{eqnarray}
  \item $\Phi(f)$ transforms covariantly under the representation
    $U$ of $\PGpo$, cf. (\ref{rep_U}):
    \begin{equation}\label{phi_cov}
      U(g)\Phi(f)U(g)^{-1} = \Phi(f_g),\quad f_g(x)=f(g^{-1}x),\quad g\in\PGpo\,.
    \end{equation}
  \item The vacuum $\Omega$ is cyclic for the field
    $\Phi$: Given any open subset $\OO\subset\Rl^2$, the
    subspace $D_\OO := \mathrm{span}\{\Phi(f_1)\cdots\Phi(f_n)\Omega
    \,:\, f_i \in \Ss(\OO),\,n\in\N_0\}$ is dense in $\Hil$.
  \item $\Phi$ is local if and only if $S_2\equiv 1$. 
  \end{enumerate}
\end{proposition}

\begin{proof}
  1) follows directly from the definition of $\Phi(f)$. To establish
  2), we note $(\overline{f})^\pm =\overline{f^\mp}$, which implies
  \begin{equation*}
  \Phi(f)^*\psi = (\zd(f^+)^* + z(f^-)^*)\psi = (z(\overline{f^+}) + \zd(\overline{f^-}))\psi = \Phi(\fbar)\psi
\end{equation*}
for $\psi\in\DD$. In particular, $\Phi(f)$ is hermitian for real $f$.\\
Now let $\psi^{(n)}\in\Hil_n$ and $N_f := \|f^+\|+\|f^-\|$. From the
definitions (\ref{a3},\ref{a4}) and $\|P_n\|=1$ we get the estimates
$\|\Phi(f)\psi^{(n)}\| \leq \sqrt{n+1}\,N_f\|\psi^{(n)}\|$ and 
\begin{equation*}
  \|\Phi(f)^k\psi^{(n)}\| \leq
  \sqrt{n+k}N_f\|\Phi(f)^{k-1}\psi^{(n)}\|\leq \sqrt{n+k}\cdots\sqrt{n+1}\, N_f^k\|\psi^{(n)}\|,\quad k\in\N.
\end{equation*}
Thus, for $\zeta\in\Cl$ we have
\begin{equation*}
  \sum_{k=0}^\infty\frac{|\zeta|^k}{k!}\|\Phi(f)^k\psi^{(n)}\| \leq \|\psi^{(n)}\|\sum_{k=0}^\infty \sqrt{\frac{(n+k)!}{n!}}\frac{1}{k!}(|\zeta|\,N_f)^k < \infty\,,
\end{equation*}
which shows that every $\psi\in\DD$ is an entire analytic vector for
$\Phi(f)$. Since $\DD$ is dense in $\Hil$, we can use Nelson's
theorem to conclude that $\Phi(f)$ is essentially selfadjoint on $\DD$. In the
following we use the same symbol $\Phi(f)$ for the selfadjoint closure
of this operator.\\
3) follows immediately from $((\Box+m^2)f)^\pm=0$. 
To prove 4) we choose $\varphi\in\Hil_1,\psi^{(n)}\in\Hil_n$,
$g\in\PGpo$ and calculate 
\begin{eqnarray*}
  U(g)\zd(\varphi)U(g)^*\psi^{(n)} &=&
  \sqrt{n+1}\;U(g)P_{n+1}(\varphi\otimes U(g)^*\psi^{(n)})\\
  &=& \sqrt{n+1}\;P_{n+1}(U(g)\varphi\otimes\psi^{(n)})\\
  &=& \zd(U(g)\varphi)\psi^{(n)}\;.
\end{eqnarray*}
This implies $U(g)z(\varphi)U(g)^* = z(J_0 U(g)J_0\varphi)$. Now let $g=(a,\la)$. Because of $J_0 U(a,\la) J_0 = U(-a,\la)$
one gets $U(g)\Phi(f)U(g)^* = \zd(U(a,\la)f^+) +
z(U(-a,\la)f^-)$, and the latter expression is equal to $\Phi(f_{(a,\la)})$ since
${f_{(a,\la)}}^\pm = U(\pm a,\la)f^\pm$.\\
5) Let $\mathscr{P}(\OO)$ denote the algebra generated by all
polynomials in the field $\Phi(f)$ with $f\in\Ss(\OO)$. By standard
analyticity arguments making use of the spectrum condition it follows that $\mathscr{P}(\OO)\Omega$ is dense in $\Hil$ if
and only if $\mathscr{P}(\Rl^2)\Omega$ is dense in $\Hil$. Choosing
$f\in\Ss(\Rl^2)$ such that $f^-=0$, we conclude that
$\zd(f^+)\in\mathscr{P}(\Rl^2)$ which implies that
$\mathscr{P}(\Rl^2)\Omega$ is dense in $\Hil$.\\
6) One cannot expect the field $\Phi(f)$ to be local since it creates one-particle states from the vacuum ($\Phi(f)\Omega=f^+\in\Hil_1$), a property which is known to be compatible with locality only in the free case, as the Jost-Schroer-Theorem states \cite{allthat}. Explicitly one may look at the two particle contribution of the field commutator applied to the vacuum:
\begin{eqnarray*}
  P_2[\Phi(f),\Phi(g)]\Omega &=& \sqrt{2}P_2(f^+\otimes g^+ - g^+\otimes f^+)\\
  &=& 2^{-1/2}(f^+\otimes g^+ -g^+\otimes f^+ - S_2^*(f^+\otimes g^+ - g^+\otimes f^+))\,.
\end{eqnarray*}
If $S_2\equiv 1$ this expression vanishes, as expected from the locality of
the free field. However, $\Phi$ is not local in the general case.
\end{proof}
{\em Remark}: It is worth mentioning that $\Phi(f)$ fulfills the temperateness
assumption with respect to Poincar\'e transformations made in
\cite{BBS}. In fact,
\begin{eqnarray}
  \|\Phi(f)U(g)\psi^{(n)}\| \leq \sqrt{n+1}N_f\|U(g)\psi^{(n)}\| = \sqrt{n+1}N_f\|\psi^{(n)}\|
\end{eqnarray}
for any $f\in\Ss(\Rl^2)$, $g\in\PGpo$, $\psi^{(n)}\in\Hil_n$.\\
\\
Since locality is one of the fundamental principles in
relativistic quantum field theory, the non-local character of our field $\Phi$ seems
to be a severe problem. But, as we will show below, the field is not
completely delocalized. It is localizable in certain unbounded wedge regions of Minkowski space which are defined as follows. A subset of $\Rl^2$ will be called wedge if it is a Poincar\'e transform of the right hand wedge
\begin{equation}
  W_R := \{x\in\Rl^2\,:\,x_1>|x_0|\}\,.
\end{equation}
The wedge $W_R$ and its causal complement $W_L=-W_R$ will be of
special significance. Since the boost transformations leave $W_R$
invariant, the set of all wedges $\W$ is given by
\begin{equation}\label{wedgeset}
  \W = \bigcup_{x\in\Rl^2}\{W_R+x,W_L+x\}\;.
\end{equation}
Starting from $\Phi$, we will exhibit for every wedge $W\in\W$ quantum
fields localized in $W$.
These fields transform covariantly under the given representation $U$ of the
Poincar\'e group, have the vacuum as a cyclic vector and commute if
they are localized in space-like separated wedges, i.e. they are of
Bose type.\\
We assign the field $\Phi(f)$ to the left wedge $W_L+x$ whenever
$\supp(f)\subset W_L+x$. In view of (\ref{phi_cov}) this assignment is
covariant with regard to the automorphic action of $\PGpo$.\\
The crucial step is to exhibit a covariant field $\Phi'$ such that
$\Phi'(f)$ is localized in $W_R+y$ if $\supp(f)\subset W_R+y$, and
which is local relative to $\Phi$.
Schroer suggested \cite{schroer} that $\Phi$ and $\Phi'$ are related
by the adjoint action of the operator 
\begin{equation}
  J:=\bigoplus_{n=0}^\infty S_nJ_0\tp{n}\quad\quad\mathrm{where}\qquad S_n := \prod_{1\leq i<j\leq n}S_n^{j,i}\quad.
\end{equation}
The properties of $J$ are specified in the following lemma.
\begin{lemma}
  \begin{enumerate}
  \item   $J$ is an antiunitary involution on $\Hil$, i.e. $J=J^{-1}=J^*$. It acts on $\Hil_n$ according to
  \begin{equation}\label{jact}
    JP_n = P_n D^+_n(\iota_n) J_0\tp{n}\;,
  \end{equation}
  where $\iota_n\in\frS_n$ denotes the permutation
  $(1,...,n)\mapsto(n,...,1)$.
\item $J$ represents the space-time reflection $j:x\longmapsto -x$,
  i.e. for $g\in\PGpo$ we have
  \begin{eqnarray}\label{j_geo}
    JU(g)J = U(jgj)\,.
  \end{eqnarray}
\end{enumerate}
\end{lemma}
\begin{proof}
1) The (anti-) unitarity of $S_n$, $J_0\tp{n}$ and (\ref{scon}) imply
that $J_n := S_nJ_0\tp{n}$ satisfies $J_n=J_n^{-1}=J_n^*$ which
entails the same properties for $J$. Thus $J$ is an antiunitary
involution on $\mathcal{F}_{\Hil_1}$.
To prove (\ref{jact}), we first note that the equality $\iota_n = \sigma_2\cdots\sigma_n$ involving the special permutations $\sigma_k$
defined in (\ref{sigma}) holds. Next we want to show $D_n(\iota_n)=S_nD_n^+(\iota_n)$ by induction in $n$, the case $n=1$ being trivial. The step from $n$ to $n+1$ is achieved with the help of (\ref{d_sigma}):
\begin{eqnarray*}
  D_{n+1}(\iota_{n+1}) &=& D_{n+1}(\iota_n\sigma_{n+1}) = (S_n D_n^+(\iota_n)\otimes 1_1)\prod_{j=1}^n S_{n+1}^{n+1,j}D_{n+1}^+(\sigma_{n+1})\\
   &=& (S_n\otimes 1_1) \prod_{j=1}^n S_{n+1}^{n+1,n+1-j}(D_n^+(\iota_n)\otimes 1_1)D_{n+1}^+(\sigma_{n+1})\\
   &=& S_{n+1}D_{n+1}^+(\iota_{n+1})\,.
\end{eqnarray*}
Since $D_n(\iota_n)=1$ on $\Hil_n$, this implies that $J$ acts on $\Hil_n$ like
$J_n=S_nJ_0\tp{n}=J_0\tp{n}S_n^*=J_0\tp{n}D_n^+(\iota_n)$. It remains
to show that $J_0\tp{n}D_n^+(\iota_n)$ commutes with $P_n$. The equations $\iota_n \tau_i \iota_n = \tau_{n-i}$ and
$\iota_n^2=\mathrm{id}$ hold in $\frS_n$ and lead to
\begin{eqnarray*}
  J_0\tp{n}D_n^+(\iota_n)D_n(\tau_i) &=& 
  J_0\tp{n}D_n^+(\iota_n)S_n^{i+1,i}D_n^+(\tau_i)\\
  &=& J_0\tp{n}S_n^{n-i,n+1-i}D_n^+(\iota_n\tau_i)\\
  &=& S_n^{n-i+1,n-i}J_0\tp{n}D_n^+(\tau_{n-i}\iota_n)\\
  &=& D_n(\tau_{n-i})\,J_0\tp{n}D_n^+(\iota_n)\;.
\end{eqnarray*}
Considering a general permutation $\pi=\tau_{i_1}\cdots\tau_{i_k}$,
this implies that $J$ acts on the representation of $\frS_n$
via $J_0\tp{n}D_n^+(\iota_n)D_n(\pi) =
D_n(\pi')J_0\tp{n}D_n^+(\iota_n)$, $\pi' := \iota_n \pi
\iota_n$. Hence the mean over the group, the symmetrization $P_n$, is
left invariant. So we conclude that (\ref{jact}) holds and $J$ can be
restricted to $\Hil$.\\
2) To prove (\ref{j_geo}), we notice that $JU_1(a,\la)J = U_1(-a,\la) =
U_1(j(a,\la) j)$ holds on the one-particle space for any
$(a,\la)\in\PGpo$. But in view of (\ref{jact}), this equality holds on
$\Hil_n$ and thus on $\Hil$, too.
\end{proof}

Using this TCP operator $J$, we define the ``reflected''
Zamolodchikov operators
\begin{eqnarray}
  \zd(\varphi)' := J\zd(\varphi)J,\qquad z(\varphi)' := Jz(\varphi)J,
\end{eqnarray}
and a ``reflected'' field
\begin{eqnarray}\label{phi'}
   \Phi'(f) := J\Phi(f^j)J\;,\qquad f^j(x):=\overline{f(-x)}\;.
\end{eqnarray}
These are linear, unbounded operators defined on $\DD$. The field
$\Phi'(f)$ has the same properties as $\Phi(f)$: It depends linearly
on $f$, is essentially self-adjoint on $\DD$ if $f$ is real and
provides a solution of the Klein-Gordon equation (Note $\Phi'(f)\Omega=f^+=\Phi(f)\Omega$). $\Phi'(f)$
transforms covariantly under the representation $U$ of $\PGpo$. In view of $J^2=1$ and $J\Omega=\Omega$ we have 
\begin{eqnarray*}
  \Phi'(f_1)\cdots\Phi'(f_n)\Omega = J\Phi(f_1^j)\cdots\Phi(f_n^j)\Omega
\end{eqnarray*}
and conclude that $\Omega$ is cyclic for $\Phi'$ as well. Furthermore,
$\Phi'(f)$ will turn out to be localized in $W_R+\supp(f)$. As a prerequisite for this important result, we
compute the commutation relations of creation and annihilation
operators with their reflected counterparts in the following lemma.
\begin{lemma}
  Let $\varphi_1,\varphi_2 \in\Hil_1$. The following commutation
  relations hold on $\DD$:
  \begin{eqnarray}
    \big[\zd(\varphi_1)',\zd(\varphi_2)\big] &=& 0,\qquad\qquad \big[ z(\varphi_1)' ,z(\varphi_2)  \big] = 0\;.
  \end{eqnarray}
  The ``mixed'' commutators act on $\Hil_n$, $n\in\N_0$, as multiplication operators ($\underline{\te}=(\te_1,...,\te_n)$):
  \begin{eqnarray*}
    \big[\zd(\varphi_1)',z(\varphi_2)\big]\psi^{(n)} \!\!&=&\!\! B_n^{\varphi_1,\varphi_2}\cdot\psi^{(n)},\;\; B_n^{\varphi_1,\varphi_2}(\underline{\te})=
    -\int\!\mathrm{d}\te\,\overline{\varphi_1(\te)}\varphi_2(\te)\prod_{j=1}^n
    S_2(\te_j-\te),\label{Bn}\\
    \big[z(\varphi_1)',\zd(\varphi_2)\big]\psi^{(n)} \!\!&=&\!\! C_n^{\varphi_1,\varphi_2}\cdot\psi^{(n)},\;\; C_n^{\varphi_1,\varphi_2}(\underline{\te})=+
    \int\mathrm{d}\te\,\,\overline{\varphi_1(\te)}\varphi_2(\te)\prod_{j=1}^n
    S_2(\te-\te_j).
  \end{eqnarray*}
\end{lemma}
\begin{proof}
  With the help of the preceeding lemma we can compute the action of
  $\zd(\varphi)'$ and $z(\varphi)'$, $\varphi\in\Hil_1$, on $\Hil_n$:
  \begin{eqnarray*}
    \zd(\varphi)'\psi^{(n)} &=& \sqrt{n+1}JP_{n+1}(\varphi\otimes
    J\psi^{(n)}) = \sqrt{n+1}P_{n+1}(\psi^{(n)}\otimes J_0 \varphi)\,,\\
    z(\varphi)'\psi^{(n)} &=& \sqrt{n}J I_{n+1}^{1,2}[\varphi\otimes
    J\psi^{(n)}] = \sqrt{n} I_{n+1}^{1,n+1}[J_0\varphi\otimes\psi^{(n)}]\quad.   \end{eqnarray*}
The creation operators $\zd(\varphi_1)$ commute with $\zd(\varphi_2)'$
on $\Hil_n$ because these operators perform the independent operations of
tensorial multiplication from the left and right, respectively. So $[\zd(\varphi_1)',\zd(\varphi_2)]\psi = 0$ for all $\varphi_1,\varphi_2\in\Hil_1$, $\psi\in\DD$, and a similar argument leads to $[z(\varphi_1)',z(\varphi_2)]\psi=0$.\\
It remains to investigate the mixed commutation relations. We use a
decomposition of permutations $\pi\in\frS_{n+1}$ similar to the one used
in the proof of Lemma \ref{z-lemma}, namely $D_{n+1}(\pi) =
D_{n+1}(\sigma_k')(D_n(\rho)\otimes 1_1)$. (In the
following the notation $\varphi^*:=J_0\varphi$ for
vectors $\varphi\in\Hil_1$ is used.)
\begin{eqnarray*}
\lefteqn{[\zd(\varphi_1)',z(\varphi_2)]\psi^{(n)} =
  \sqrt{n}P_n(z(\varphi_2)\psi^{(n)}\otimes \varphi_1^*)
 - \sqrt{n+1}I_{n+2}^{1,2}[\varphi_2\otimes \zd(\varphi_1)'\psi^{(n)}]}\\
  &=& nP_n I_{n+2}^{1,2}[\varphi_2\otimes\psi^{(n)}\otimes
  \varphi_1^*] - (n+1)I_{n+2}^{1,2}[\varphi_2\otimes
  P_{n+1}(\psi^{(n)}\otimes \varphi_1^*)]\\
  &=& n I_{n+2}^{1,2}[(1_2\otimes 
  P_n)(\varphi_2\otimes\psi^{(n)}\otimes \varphi_1^*)] -
  I_{n+2}^{1,2}\big[\varphi_2\otimes\sum_{k=1}^{n+1}D_{n+1}(\sigma_k')(\psi^{(n)}\otimes \varphi_1^*)\big]
\end{eqnarray*}
This expression vanishes apart from the term corresponding to $k=n+1$, as
can be seen using $D_{n+1}(\sigma_k') = 1_1\otimes D_n(\sigma_k')$, $k<n+1$, and the decomposition $P_n=\frac{1}{n}\sum_{r=1}^nD_n(\sigma_r')(P_{n-1}\otimes
1_1)$.
Taking into account $\sigma_{n+1}'=\tau_1\cdots\tau_n$ and
$$
D_{n+1}(\sigma_{n+1}') = S_{n+1}^{2,1}D_{n+1}^+(\tau_1)\cdots
S_{n+1}^{n+1,n}D_{n+1}^+(\tau_n) = \prod_{j=2}^{n+1}S_{n+1}^{j,1}
D_{n+1}^+(\sigma_{n+1}')\,,
$$
we have thus shown
 \begin{eqnarray*}
   [\zd(\varphi_1)',z(\varphi_2)]\psi^{(n)} &=& 
   -I_{n+2}^{1,2}\left[\varphi_2\otimes
     D_{n+1}(\sigma_{n+1}')(\psi^{(n)}\otimes \varphi_1^*)\right]\\
   &=& -I_{n+2}^{1,2}\bigg[ \varphi_2\otimes \prod_{j=1}^n
      S_{n+1}^{1+j,1}\varphi_1^*\otimes \psi^{(n)}\bigg]\;.
 \end{eqnarray*}
So this commutator acts on $\Hil_n$ by multiplication with the function $B_n^{\varphi_1,\varphi_2}$. As
\begin{eqnarray*}
  [z(\varphi_1)',\zd(\varphi_2)]\psi^{(n)} &=& -[\zd(\varphi_1^*)',z(\varphi_2^*)]^*\psi^{(n)} = -\overline{B_n^{\varphi_1^*,\varphi_2^*}}\cdot\psi^{(n)}\,,
\end{eqnarray*}
and $\overline{S_2(\te_j-\te)}=S_2(\te-\te_j)$, we also obtain $C_n^{\varphi_1,\varphi_2}= -\overline{B_n^{\varphi_1^*,\varphi_2^*}}$ as claimed.
\end{proof}

Let us now assume that $S_2$ is analytic on the open strip $S(0,\pi)$,
and bounded on its closure. Since poles in the scattering function
indicate bound states \cite{abda}, we exclude hereby the possibility of bound
states in agreement with the formulation of the model describing a
single species of particles.\\
We will show now that the field $\Phi'(f)$ is localized in
$W_R+\supp(f)$. Since we assigned $\Phi(g)$ to the wedge $W_L+x$ if
$\supp(g)\subset W_L+x$, we have to show that $\Phi'(f)$ and $\Phi(g)$
commute in the sense that $[\Phi'(f),\Phi(g)]$ vanishes on $\DD$ if $W_L+\supp(g)$ is spacelike to $W_R+\supp(f)$. This is the
case if and only if $\supp(f)-\supp(g)\subset W_R$. Using the
translation covariance of $\Phi$ and $\Phi'$, it suffices to consider
the case $\supp(f)\subset W_R$, $\supp(g)\subset W_L$ only.\\
Let $f\in\Ss(W_R)$, $g\in\Ss(W_L)$. Since $g$ has support in the cone
$W_L$, its Fourier transform $\gti$ is the boundary value of a function analytic
in the tube $\Rl^2 - i W_L$ \cite{reed2}. Using the rapidity parametrization (\ref{Phi}) one calculates that 
\begin{equation*}
  \mathrm{Im}(p(\te + i\mu)) = m\sin\mu\left(
    \begin{array}{c}
      \sinh\te\\\cosh\te
    \end{array}
\right) \in -W_L\;\;\mathrm{for}\; 0<\mu<\pi\;.
\end{equation*}
Thus $g^+$ has an analytic continuation to $S(0,\pi)$. Since
$p(\te+i\pi) = -p(\te)$, the value at the upper boundary of
the strip is given by $g^+(\te+i\pi) = g^-(\te)$. Considering $W_R$
instead of $W_L$ one concludes that $g^-$ as well as $\overline{g^+}$
are analytic on the strip $S(-\pi,0)$. The fact
that $g$ is a Schwartz function and
$|e^{ip(\te)x}|=e^{-\mathrm{Im}(p(\te))\cdot x}$, $\mathrm{Im}(p(\te))x
\geq 0$ for $\mathrm{Im}(\te)\in [0,\pi]$ and $x\in W_L$, imply that
$g^+(\te)$, $\mathrm{Im}(\te)\in (0,\pi)$ decays exponentially to zero for
$\mathrm{Re}(\te) \to \pm\infty.$ On the boundary we have $g^+(\te)\to
0$, $g^-(\te) = g^+(\te+i\pi)\to 0$ for $\te\to\pm\infty$ since $g$ is a
Schwartz space test function.\\
All these considerations apply to $f^j$
as well since $f^j\in\Ss(W_L)$. In view of
$(f^j)^\pm=\overline{f^\pm}$ we have analyticity for $f^+$ and $f^-$ in
$S(-\pi,0)$ and $S(0,\pi)$, respectively, as well as exponential decay
for $\mathrm{Re}(\te)\to\pm\infty$ in these strips.\\
Having recalled these analyticity properties we now want to prove that
$\Phi'(f)$ and $\Phi(g)$ commute on $\Hil_n$. In view of the
preceding lemma the commutator simplifies to $[\Phi'(f),\Phi(g)] = [\zd(\overline{f^+})',z(g^-)] +
[z(\overline{f^-})',\zd(g^+)]$ and it remains to show that
$B_n^{\overline{f^+},g^-} + C_n^{\overline{f^-},g^+} = 0$. The integrand of $C_n^{\overline{f^-},g^+}$ is analytic on $S(0,\pi)$ and of fast decrease for
$\mathrm{Re}(\te)\to\pm\infty$ because the test functions decay
exponentially and $S_2$ is bounded on $S(0,\pi)$. This enables us to
shift the integration from $\Rl$ to $\Rl+i\pi$:
\begin{eqnarray*}
  C_n^{\overline{f^-},g^+}(\te_1,...,\te_n) &=& \int\mathrm{d}\te\,
  f^-(\te)g^+(\te)\prod_{j=1}^n S_2(\te-\te_j)\\
  &=& \int\mathrm{d}\te\, f^-(\te+i\pi)g^+(\te+i\pi)\prod_{j=1}^n S_2(\te+i\pi -\te_j)\\
  &=& \int\mathrm{d}\te\, f^+(\te)g^-(\te)\prod_{j=1}^n
  S_2(\te_j-\te)\\
  &=& - B_n^{\overline{f^+},g^-}(\te_1,...,\te_n)\;,
\end{eqnarray*}
where in the third equality we used the crossing symmetry $S_2(\te+i\pi)=S_2(-\te)$. Hence $[\Phi'(f),\Phi(g)]$
vanishes on $\DD$ if $S_2$ is analytic in $S(0,\pi)$.\\
Since we have the vacuum as an analytic
vector for $\Phi'(f)$ and $\Phi(g)$, we may apply an argument of
Borchers and Zimmermann \cite{bozi} to conclude that the unitary
groups $e^{it\Phi'(f)}$ and $e^{is\Phi(g)}$, $f,g\in\Ss(W_R)$ real,
$t,s\in\Rl$, commute as well. Thus one can proceed from the fields to
nets of local von Neumann algebras affiliated with wedges.\\
We summarize our results in the following proposition:
\begin{proposition} Let $S_2$ be a bounded, continuous function on
  $\overline{S(0,\pi)}$ which is analytic on the open strip $S(0,\pi)$
  and satisfies the constraints (\ref{s_constraints}).
  \begin{enumerate}
  \item The field $\Phi'$ has the properties 1)-6) stated in
    Proposition 1 for the field $\Phi$.
  \item Let $f, g\in\Ss(\Rl^2)$. If $W_R+\supp(f)$ and $W_L+\supp(g)$
    lie spacelike to each other, $\Phi'(f)$ and $\Phi(g)$ commute. 
  \end{enumerate}
\end{proposition}
The assignment of $\Phi'(f)$ to the wedge $W_L+x$ whenever $\supp(f)\subset
W_L+x$ thus completes the definition of our wedge-local quantum field theory.\\
It remains to be shown that the scattering of our model is given by
$S_2$. This will be done in the next section. Due to the
weak localization properties of $\Phi$ and $\Phi'$, we have to restrict
our analysis to collision processes of two particles.

\section{Scattering States}
\setcounter{equation}{0}
For the analysis of scattering processes we employ the Haag-Ruelle
scattering theory in a form used in \cite{BBS} for wedge-localized
fields. We briefly recall how two-particle scattering states can be
constructed in this case.\\
We first introduce some notation: The velocity support of a
test function $f$ will be denoted by
$\Gamma(f):=\{(1,p_1/\omega_p)\in\Rl^2\,:\,p\in\supp\fti\}$, where $\omega_p=\sqrt{p_1^2+m^2}$, and we write
$\Gamma(g)\prec\Gamma(f)$ if $\Gamma(f)-\Gamma(g)\subset W_R$. One
verifies that $\Gamma(g)\prec\Gamma(f)$ implies that the support of
$g^+$ lies left to the support of $f^+$, i.e. $\supp(f^+) -\supp(g^+) \subset\R^+$.\\
Furthermore, we use the family of functions depending on the time
parameter $t$
\begin{equation}
  f_t (x) := \frac{1}{2\pi}\int\!\mathrm{d}^2p\,\tilde{f}(p)
  e^{i(p_0- \omega_p) t}e^{-ipx}\label{f_t}\,.
\end{equation}
It is well-known that the support of the function $f_t$ is essentially
localized in $t\,\Gamma(f)$ for large $|t|$.\\ 
To construct outgoing scattering states, we consider test functions $f$ and $g$
with $\Gamma(g)\prec\Gamma(f)$ and whose energy-momentum supports are
contained in a neighbourhood of the mass shell $H^+_m=\{(p_1,\sqrt{p_1^2+m^2}):p_1\in\Rl^2\}$. Thus at very late or early
times, the operators $\Phi'(f)$ and $\Phi(g)$ are essentially
localized in $W_R+t\,\Gamma(f)$ and $W_L+t\,\Gamma(g)$,
respectively \cite{BBS}. Since $\Gamma(g)\prec\Gamma(f)$, these
localization regions are spacelike separated for large positive $t$
and their distance increases linearly with $t$. In view of
$\Phi(g)\Omega = g^+$, $\Phi'(f)\Omega=f^+$, one can therefore define
the outgoing two-particle state
\begin{eqnarray}
\;\;\;\;\;\,\lim_{t\to\infty}\Phi(g_t)\Phi'(f_t)\Omega = (g^+ \times f^+)\oout\,,
\end{eqnarray}
and similarly 
\begin{equation}
\lim_{t\to\infty}\Phi'(f_t)\Phi(g_t)\Omega = (f^+\times g^+)\oout\,.
\end{equation}
Since the operators $\Phi'(f_t)$ and $\Phi(g_t)$ commute for
$t\to\infty$, we have symmetric scattering states $ (g^+\times f^+)\oout= (f^+\times g^+)\oout$ as required for a Boson.\\
To construct incoming scattering states one has to exchange $f$ and
$g$ because then $W_L+t\,\Gamma(f)$ and $W_R+t\,\Gamma(g)$ are far apart and
spacelike separated in the limit $t\to-\infty$. Hence
\begin{eqnarray}
\lim_{t\to-\infty}\Phi(f_t)\Phi'(g_t)\Omega = (f^+\times g^+)\iin =
(g^+\times f^+)\iin\,.
\end{eqnarray}
In our model these limits can be easily computed: The definition
(\ref{f_t}) implies ${f_t}^+=f^+$, ${g_t}^+=g^+$ and furthermore we
have ${f_t}^-=0$, ${g_t}^-=0$ since the supports of $\fti$, $\gti$ do not intersect the lower mass shell $-H_m^+$.\\
So the time-dependence drops out and we arrive at the
following form of the two-particle scattering states:
\begin{eqnarray*}
  (g^+\times f^+)\oout &=& \lim_{t\to\infty}\Phi(g_t)\Phi'(f_t)\Omega
  \;\;= \zd(g^+)f^+ = \sqrt{2}P_2(g^+ \otimes f^+),\\
  (g^+ \times f^+)\iin &=& \lim_{t\to-\infty}\Phi(f_t)\Phi'(g_t)\Omega =
  \zd(f^+)g^+ = \sqrt{2}P_2(f^+\otimes g^+)\,.
\end{eqnarray*}
Varying $f,g$ within the above limitations, we obtain a total set of
two-particle scattering states.\\
In an integrable quantum field theory in two dimensions, the
two-particle S-matrix element ${}_{\mathrm{out}}\langle
\te_1,\te_2\,|\,\te_1,\te_2\rangle\iin = S_2(|\te_1-\te_2|)$ depends
on the absolute value of the difference of the incoming rapidities
only and allows for a meromorphic continuation when restricted to
$\te_1-\te_2>0$. This meromorphic function is the scattering function
$S_2$ we have worked with so far. The two-particle S-matrix, however,
acts on $\Hil_2$ as the multiplication operator
\begin{eqnarray*}
  (S\psi)(\te_1,\te_2) := S_2(|\te_1-\te_2|)\psi(\te_1,\te_2)\,.
\end{eqnarray*}
Since $\supp(f^+)-\supp(g^+)\subset\Rl_+$ and $S_2(-\te)=S_2(\te)^{-1}$, we have
\begin{eqnarray*}
  S(f^+\otimes g^+) = S_2(f^+\otimes g^+)\,,\qquad\quad S(g^+\otimes
  f^+)=S_2^{-1}(g^+\otimes f^+)
\end{eqnarray*}
and
\begin{eqnarray*}
  S(g^+\times f^+)\oout &=& 2^{-1/2}S(g^+\otimes f^+ +
  S_2^*(f^+\otimes g^+))\\
  &=& 2^{-1/2}(S_2^{-1}(g^+\otimes f^+) + f^+\otimes g^+)\\
  &=& \sqrt{2}P_2(f^+\otimes g^+)\\
  &=& (g^+\times f^+)\iin\,.
\end{eqnarray*}
Having verified the action of $S$ on a total set of scattering
states, it follows by linearity and continuity that the two-particle S-matrix of our model is the
multiplication operator $S$ corresponding to the scattering function $S_2$.\\
\\
We have shown that to many factorizing S-matrices a
wedge-local theory with the correct scattering behaviour can be constructed
with the help of polarization-free generators. Of course the important
question arises wether there exist local fields in our model and how
they are related to the S-matrix. As Schroer pointed out, this
question amounts to the analysis of intersections of certain operator
algebras \cite{schroer} which is another crucial step in the approach to the inverse scattering problem presented here.


\subsection*{Acknowledgements}
I would like to thank D.~Buchholz for many discussions which
have been essential for the work. Special thanks are also due to
J. Mund for helpful conversations and valuable advice. Financial
support by the Lucia-Pfohe Stiftung is gratefully acknowledged.


\end{document}